\documentclass[review,12pt,authoryear]{elsarticle}

\usepackage{amssymb}
\usepackage{amsthm}
\usepackage{framed} 
\usepackage{amsmath,color}
\usepackage{mathrsfs}
\usepackage{graphicx}
\usepackage{epstopdf}
\usepackage{float}
\usepackage{caption}
\usepackage{subcaption}
\usepackage{bm}
\usepackage{bbm}
\usepackage{mathrsfs}
\usepackage{cleveref}
\usepackage{soul}
\usepackage{accents}
\usepackage{color,soul} 
\usepackage{color} 
\usepackage{nomencl} 
\biboptions{sort&compress}
\soulregister\citep7 
\soulregister\citet7 
\soulregister\citealp7 
\newsavebox{\measurebox} 
\usepackage{titlesec} 
\usepackage{tabu} 
\usepackage{longtable}

\journal{European Journal of Mechanics A/Solids}

\makeatletter
\def\@author#1{\g@addto@macro\elsauthors{\normalsize%
    \def\baselinestretch{1}%
    \upshape\authorsep#1\unskip\textsuperscript{%
      \ifx\@fnmark\@empty\else\unskip\sep\@fnmark\let\sep=,\fi
      \ifx\@corref\@empty\else\unskip\sep\@corref\let\sep=,\fi
      }%
    \def\authorsep{\unskip,\space}%
    \global\let\@fnmark\@empty
    \global\let\@corref\@empty  
    \global\let\sep\@empty}%
    \@eadauthor={#1}
}
\makeatother

\titleformat{\paragraph}
{\normalfont\normalsize\itshape}{\theparagraph}{1em}{}
\titlespacing*{\paragraph}
{0pt}{3.25ex plus 1ex minus .2ex}{1.5ex plus .2ex}

\begin{document}

\begin{frontmatter}



\title{Gradient-enhanced statistical analysis of cleavage fracture}


\author{Emilio Mart\'{\i}nez-Pa\~neda\corref{cor1}\fnref{Cam}}
\ead{mail@empaneda.com}

\author{Sandra Fuentes-Alonso\fnref{Uniovi}}

\author{Covadonga Beteg\'{o}n\fnref{Uniovi}}

\address[Cam]{Department of Engineering, Cambridge University, CB2 1PZ Cambridge, UK}

\address[Uniovi]{Department of Construction and Manufacturing Engineering, University of Oviedo, Gij\'{o}n 33203, Spain}

\cortext[cor1]{Corresponding author.}

\begin{abstract}
We present a probabilistic framework for brittle fracture that builds upon Weibull statistics and strain gradient plasticity. The constitutive response is given by the mechanism-based strain gradient plasticity theory, aiming to accurately characterize crack tip stresses by accounting for the role of plastic strain gradients in elevating local strengthening ahead of cracks. It is shown that gradients of plastic strain elevate the Weibull stress and the probability of failure for a given choice of the threshold stress and the Weibull parameters. The statistical framework presented is used to estimate failure probabilities across temperatures in ferritic steels. The framework has the capability to estimate the three statistical parameters present in the Weibull-type model without any prior assumptions. The calibration against experimental data shows important differences in the values obtained for strain gradient plasticity and conventional J2 plasticity. Moreover, local probability maps show that potential damage initiation sites are much closer to the crack tip in the case of gradient-enhanced plasticity. Finally, the fracture response across the ductile-to-brittle regime is investigated by computing the cleavage resistance curves with increasing temperature. Gradient plasticity predictions appear to show a better agreement with the experiments.
\end{abstract}


\end{frontmatter}



\section{Introduction}
\label{Sec:Introduction}

Macroscopic fracturing in metallic materials depends sensitively on properties that pertain to the micro and atomic scales. Not surprisingly, a considerable effort has been made to link scales in fracture mechanics, with the ultimate goal of quantitatively predicting the strength, durability, and reliability of structural components (\citealp{Suo1993}; \citealp{Hutchinson1997}). These
endeavours aim at spanning the wide range of scales at stake by enriching continuum theories to properly characterize behaviour at the small scales involved in crack tip deformation.\\

The deficiencies intrinsic to conventional plasticity theory provide a strong motivation for developing mechanistically-based models. Namely, unrealistically low stresses are predicted ahead of the crack tip, with toughness being unbounded for cohesive strengths of approximately 3 times the yield stress in a perfectly plastic material ($\hat{\sigma}/\sigma_Y \to 4$ in a mild hardened solid, see \citealp{Tvergaard1992}). Opening stresses on the order of 3-5 times the initial tensile yield stress fail to explain decohesion at the atomic scale. Cleavage fracture in the presence of significant plastic flow has been experimentally observed in numerous material systems (\citealp{Elssner1994}; \citealp{Bagchi1996}; \citealp{Korn2002}). Since atomic separation requires traction levels on the order of the theoretical lattice strength (10$\sigma_Y$ or larger), classic continuum theories would appear to rule out a fracture mechanism based on atomic decohesion whenever plasticity develops in the vicinity of the crack. Moreover, conventional plasticity predictions reveal important discrepancies with separation strengths calculated from first principles \citep{Raynolds1996}, and toughness bounds attained by discrete dislocation dynamics \citep{Cleveringa2000,Irani2017}, highlighting the need to bridge the gap between macroscopic modelling of cracking and the microstructural and atomistic mechanisms of fracture.\\

Small scale experiments have consistently shown that conventional plasticity theory is unable to characterize the material response of metals at the micro level. Fostered by growing interest in microtechnology, a wide range of mechanical tests on micro-sized specimens have revealed that metallic materials display strong size effects when deformed non-uniformly into the plastic range. Experiments such as indentation \citep{Nix1998}, torsion \citep{Fleck1994}, or bending \citep{Stolken1998} predict a 3-fold increase in the effective flow stress by reducing specimen size (\emph{smaller is stronger}). This size effect is attributed to gradients of plastic strain that require a definite density of dislocations to accommodate lattice curvature \citep{Ashby1970}. These geometrically necessary dislocations (GNDs) are not accounted for in conventional theories of plasticity, neglecting the length scale dependency intrinsically associated with plastic flow. A large theoretical literature has appeared seeking to extend plasticity theory to small scales by the development of isotropic strain gradient plasticity (SGP) formulations (\citealp{Aifantis1984}; \citealp{Gao1999}; \citealp{Fleck2001}; \citealp{Anand2005}). Using SGP theories to provide an implicit multi-scale characterization of the mechanical response ahead of a crack appears imperative as, independently of the size of the specimen, the plastic zone adjacent to the crack tip is physically small and contains strong spatial gradients of deformation \citep{IJSS2015}. The investigation of stationary crack tip fields has shown that plastic strain gradients promote local strain hardening and lead to much higher stresses relative to classic plasticity predictions (\citealp{Jiang2001}; \citealp{Wei2006}; \citealp{Komaragiri2008}; \citealp{CM2017}). Accurately capturing crack tip stresses has proven to be fundamental in predicting fatigue damage (\citealp{Sevillano2001}; \citealp{Brinckmann2008}; \citealp{Pribe2019}), notch fracture mechanics \citep{TAFM2017}, microvoid cracking \citep{Tvergaard2008}, and hydrogen assisted failure \citep{IJHE2016,AM2016}. Since plastic strain gradients can alter crack tip stresses over several tens of $\mu$m, it is expected that strain gradient plasticity models will also play an important role in the modelling of cleavage fracture and the ductile-to-brittle transition \citep{Qian2011,JMPS2019}.\\

Cleavage fracture models are grounded on the concept of microcracks nucleating from defects, such as inclusions or second-phase particles \citep{Pineau2016}. The location of these defects is statistical by nature and, consequently, modelling efforts rely mainly on probabilistic analysis. The seminal work by the Beremin group \citep{Beremin1983} established the fundamental framework on which most cleavage models stand; Weibull statistics and the weakest link model are employed to estimate the probability of failure $P_f$, where $P_f$ equals the probability  of sampling (at least) one critical fracture-triggering particle. In these models the stress level is the driving force for fracture and, consequently, local strengthening due to plastic strain gradients will influence failure probability predictions.\\

In this work we make use of a mechanism-based strain gradient plasticity formulation to accurately characterize crack tip stresses. The constitutive description is coupled with a probabilistic framework capable of obtaining all the statistical parameters of the model without any prior assumptions. The capabilities of the present mechanism-based scheme for probabilistic analysis of brittle fracture are benchmarked against experimental data from the Euro toughness project \citep{Heerens2002}. Experiments are reproduced over a wide range of temperatures, so as to span the ductile-to-brittle regime. Strain gradient plasticity predictions are compared with results from conventional plasticity and insight is gained into the role of the stress elevation due to strain gradients in assessing cleavage. 

\section{Numerical model}
\label{Sec:NumModel}

The implicitly multi-scale statistical framework for brittle fracture presented stands on a Taylor-based strain gradient plasticity formulation (Section \ref{Sec:Theory}), and a three-parametric Weibull type statistical model (Section \ref{Sec:Weibull}). The implementation is carried out by coupling a general purpose finite element program with the statistical tools of Matlab, see Section \ref{Sec:ABAQUS}. A experimental campaign employing Compact Tension specimens will be reproduced to highlight the capabilities of the model (Section \ref{Sec:CT}).

\subsection{Mechanism-based Strain Gradient Plasticity}
\label{Sec:Theory}

We model strain gradient effects by means of the so-called mechanism-based strain gradient (MSG) plasticity theory \citep{Gao1999,Qiu2003}. MSG plasticity is grounded on Taylor's dislocation model. Accordingly, the shear flow stress $\tau$ is formulated in terms of the dislocation density $\rho$ as
\begin{equation}\label{Eq1MSG}
\tau = \alpha \mu b \sqrt{\rho}
\end{equation}

\noindent where $\mu$ is the shear modulus, $b$ is the magnitude of the Burgers vector and $\alpha$ is an empirical coefficient that is taken to be equal to 0.5. The dislocation density is additively composed of the density $\rho_S$ for statistically stored dislocations (SSDs) and the density $\rho_G$ for geometrically necessary dislocations (GNDs),
\begin{equation}\label{Eq2MSG}
\rho = \rho_S + \rho_G
\end{equation}

The GND density $\rho_G$ is related to the effective plastic strain gradient $\eta^{p}$ by
\begin{equation}\label{Eq3MSG}
\rho_G = \overline{r}\frac{\eta^{p}}{b}
\end{equation}

\noindent where $\overline{r}$ is the Nye-factor which is assumed to be 1.90 for face-centered-cubic (fcc) polycrystals. \citet{Gao1999} used three quadratic invariants of the plastic strain gradient tensor to represent the effective plastic strain gradient $\eta^{p}$ as
\begin{equation}
\eta^{p}=\sqrt{c_1 \eta^{p}_{iik} \eta^{p}_{jjk} + c_2 \eta^{p}_{ijk} \eta^{p}_{ijk} + c_3 \eta^{p}_{ijk} \eta^{p}_{kji}}
\end{equation}

The coefficients were determined to be equal to $c_1=0$, $c_2=1/4$ and $c_3=0$ from three dislocation models for bending, torsion and void growth. Accordingly,
\begin{equation}
\eta^{p}=\sqrt{\frac{1}{4}\eta^{p}_{ijk} \eta^{p}_{ijk}}
\end{equation}

\noindent where the components of the strain gradient tensor are obtained from
\begin{equation}
\eta^{p}_{ijk}= \varepsilon^{p}_{ik,j}+\varepsilon^{p}_{jk,i}-\varepsilon^{p}_{ij,k}
\end{equation}

The tensile flow stress $\sigma_{flow}$ is related to the shear flow stress $\tau$ by
\begin{equation}\label{Eq4MSG}
\sigma_{flow} =M\tau
\end{equation}

\noindent with $M$ denoting the Taylor factor, which equals 3.06 for fcc metals. Rearranging (\ref{Eq1MSG}-\ref{Eq3MSG}) and substituting into (\ref{Eq4MSG}) renders
\begin{equation}\label{Eq5MSG}
\sigma_{flow} =M\alpha \mu b \sqrt{\rho_{S}+\overline{r}\frac{\eta^{p}}{b}}
\end{equation}

The SSD density $\rho_{S}$ can be readily determined from (\ref{Eq5MSG}) knowing the relation in uniaxial tension between the flow stress and the material stress-strain curve,
\begin{equation}\label{Eq6MSG}
\rho_{S} = [\sigma_{ref}f(\varepsilon^{p})/(M\alpha \mu b)]^2
\end{equation}

\noindent Here, $\sigma_{ref}$ is a reference stress and $f$ is a non-dimensional function of the plastic strain $\varepsilon^{p}$, as given by the uniaxial stress-strain curve. Substituting into (\ref{Eq5MSG}), the flow stress $\sigma_{flow}$ reads
\begin{equation}\label{EqSflow}
\sigma_{flow} =\sigma_{ref} \sqrt{f^2(\varepsilon^{p})+\ell\eta^{p}}
\end{equation}

\noindent where $\ell$ is the intrinsic material length parameter that enters the constitutive equation for dimensional consistency. The value of $\ell$ can be obtained by fitting micro-scale experiments and typically ranges between 1 and 10 $\mu$m. The model recovers the conventional plasticity solution when $\ell=0$. 

\subsection{Weibull three-parametric}
\label{Sec:Weibull}

We present a statistical framework that has the capability of predicting brittle and ductile failure and requires no prior assumptions \citep{Muniz-Calvente2015,AES2017}. First, for a given Weibull stress $\sigma_w$ and a threshold stress for crack growth $\sigma_{th}$, the cumulative probability of failure $P_f$ is given by
\begin{equation}\label{Eq:Pf}
P_f = 1 - \exp \left[ - \left( \frac{\sigma_w - \sigma_{th}}{\sigma_u} \right)^m \right]
\end{equation}

\noindent where $\sigma_u$ and $m$ respectively denote the scaling parameter and the modulus. Equation (\ref{Eq:Pf}) is defined in \citep{Beremin1983} without $\sigma_{th}$ but stresses smaller than the yield stress were considered innocuous, implying the assumption of $\sigma_{th}=\sigma_Y$. A global Weibull stress is defined based on weakest link considerations
\begin{equation}\label{Eq:Sw}
\sigma_w = \sigma_{th} + \left[ \sum_{i=1}^{n_e} \left( \sigma_1^i - \sigma_{th} \right)^m \left(V_i / V_0 \right) \right]^{(1/m)}
\end{equation}
 
\begin{figure}[H]
  \makebox[\textwidth][c]{\includegraphics[width=0.7\textwidth]{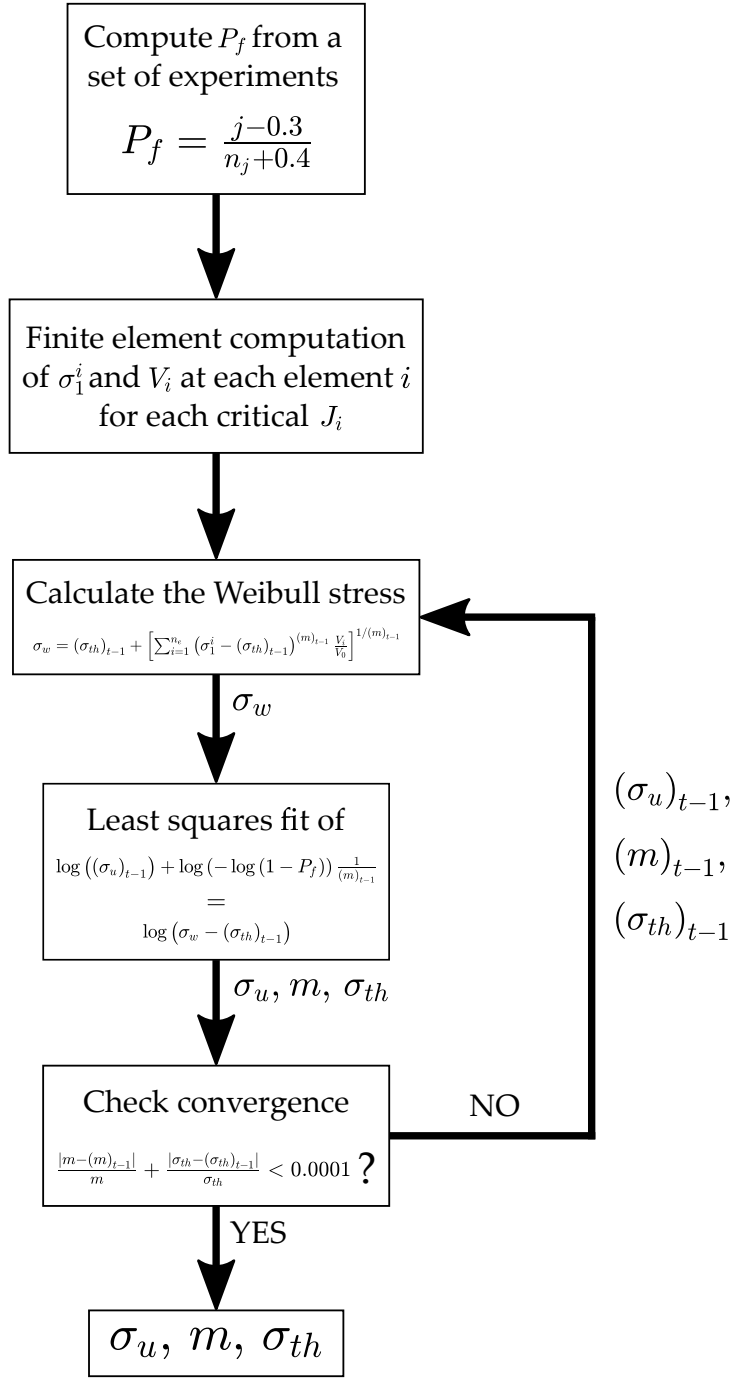}}%
  \caption{Flowchart describing the combined experimental-computational-statistical procedure for estimating the three Weibull parameters $\sigma_{th}$, $m$, and $\sigma_{th}$.}
  \label{fig:Flowchart}
\end{figure}

\noindent Here $V_0$ is a reference volume, $V_i$ is the volume of the i\emph{th} material unit in the fracture process zone experiencing a maximum principal stress $\sigma_1^i$, and $n_e$ is the number of finite elements/material units in the fracture process zone. The parameter $\sigma_{th}$ is needed due to the fact that cracks do not propagate below a certain threshold energy value. However, the concurrent estimation of the threshold, modulus and shape parameters remains a complicated task; a common approach lies in assuming a value for $\sigma_{th}$ and estimating $m$ and $\sigma_u$ from a set of experiments by means of the maximum likelihood method \citep{Muniz-Calvente2016a}. Here, all three parameters ($\sigma_{th}$, $m$ and $\sigma_u$) will be obtained by means of the following procedure: (see Fig. \ref{fig:Flowchart})\\

\noindent 1) First, the probability of failure is computed for all the experiments conducted at a given temperature. The $P_f$ versus load curve, where the load is expressed in terms of the $J$-integral, is computed by means of
\begin{equation}\label{Eq:Pfnj}
P_f = \frac{j-0.3}{n_j + 0.4}
\end{equation}

\noindent where $n_j$ denotes the number of experiments for a given temperature and $j$ is the rank number.\\

\noindent 2) A finite element analysis is conducted, and the values of $\sigma_1^i$ and $V_i$ are computed at each element for the set of critical $J_i$ at which failure has been reported in the experiments. The domain integral method is used to compute the value of $J_i$ in each load increment.\\ 

\noindent 3) The least squares method is employed to fit the Weibull distribution by using cumulative probabilities. Since the threshold parameter $\sigma_{th}$ is also an unknown, the procedure requires iterating over the following steps:\\

3.1) The Weibull stress $\sigma_w$ is first computed according to (\ref{Eq:Sw}) from the information provided by the finite element model (Step 2). In (\ref{Eq:Sw}), $m$ and $\sigma_{th}$ correspond to the values of the previous iteration (or an initial estimate, in the case of the first iteration).\\

3.2) The Weibull stress $\sigma_w$ is introduced in (\ref{Eq:Pf}) and the values of $\sigma_u$, $m$ and $\sigma_{th}$ in the current iteration are computed by fitting a univariate distribution using least squares estimates of the cumulative distribution functions. Namely, the cumulative probability of failure (\ref{Eq:Pf}) is written as,
\begin{equation}
\log \left(\sigma_u \right) + \log \left( - \log \left( 1 - P_f \right) \right) \frac{1}{m} = \log \left( \sigma_w - \sigma_{th} \right)
\end{equation}
\noindent introducing a linear relationship between $\log \left( - \log \left( 1 - P_f \right) \right)$ and $\log \left( \sigma_w - \sigma_{th} \right)$. From the $P_f$ assigned to each load (Step 1) and the Weibull stress computed for each of those loads (Step 2), we make use of least squares to fit this straight line on the transformed scale. The slope and intercept of the line provide with the values of $m$ and $\sigma_u$ for a given $\sigma_{th}$. The quality of the fit will be given by the choice of $\sigma_{th}$; we find the optimum by maximizing the coefficient of determination $R^2$ over all possible threshold values. The optimum value of $\sigma_{th}$ is specific to the current iteration and associated values of $m$ and $\sigma_u$.\\

3.3) The procedure is repeated until convergence is achieved. We assume that the solution has converged when the following criterion has been met
\begin{equation}
\frac{|\left(m\right)_t-\left(m\right)_{t-1}|}{\left(m\right)_t} + \frac{|\left(\sigma_{th}\right)_{t}-\left(\sigma_{th}\right)_{t-1}|}{\left(\sigma_{th}\right)_t} < 0.0001
\end{equation}

\noindent where $\left(m\right)_t$ denotes the value of $m$ in the current increment while the subscript $t-1$ represents its value in the previous increment. Consequently, the outcome of the analysis is the threshold value below which cracking will not occur $\sigma_{th}$, along with the two Weibull parameters $m$ and $\sigma_u$. A stress level of $\sigma_{th}+\sigma_u$ will denote a 63\% failure probability in a given material unit element.\\

\subsection{Numerical implementation}
\label{Sec:ABAQUS}

The framework presented in Sections \ref{Sec:Theory} and \ref{Sec:Weibull} is numerically implemented by exploiting Abaqus2Matlab \citep{AES2017}. Hence, we run the commercial finite element package Abaqus within the mathematical software Matlab to take advantage of Matlab's in-built capabilities for fitting univariate distributions by means of the least squares method.\\

We implement MSG plasticity in the commercial finite element package Abaqus by means of a user material subroutine (UMAT). For numerical reasons, we make use of the lower order version of MSG plasticity, commonly referred as the conventional mechanism based strain gradient (CMSG) plasticity theory \citep{Huang2004a}. As shown in \citep{IJP2016} and discussed in \citep{Shi2001}, the lower and higher order versions of MSG plasticity predict identical results except for a boundary layer of size roughly 10 nm. This boundary layer falls outside of the domain of physical validity of continuum theories; strain gradient plasticity models a collective behaviour of dislocations and it is therefore applicable at a scale much larger than the dislocation spacing. Fortran modules are used to store the plastic strain components across Gauss integration points, and the plastic strain gradient is computed by numerical differentiation within the element. First, the plastic strain increment is interpolated through its values at the Gauss points in the isoparametric space, and afterwards the increment in the plastic strain gradient is calculated by differentiation of the shape functions. The reader is referred to \citep{TAFM2017} for more details.

\subsection{Boundary value problem}
\label{Sec:CT}

We employ our framework to assess brittle failure in ferritic steels. Numerical predictions are compared to experimental results from the Euro toughness project on DIN 22NiMoCr37 steel \citep{Heerens2002}. The Euro toughness project is frequently chosen as paradigmatic benchmark for cleavage models due to the richness of its data set. Experiments are conducted at 7 temperatures, from $-154^{\circ}$C to $20^{\circ}$C, spanning the entire transition from brittle to ductile fracture.\\

\begin{figure}[H]
  \makebox[\textwidth][c]{\includegraphics[width=1.1\textwidth]{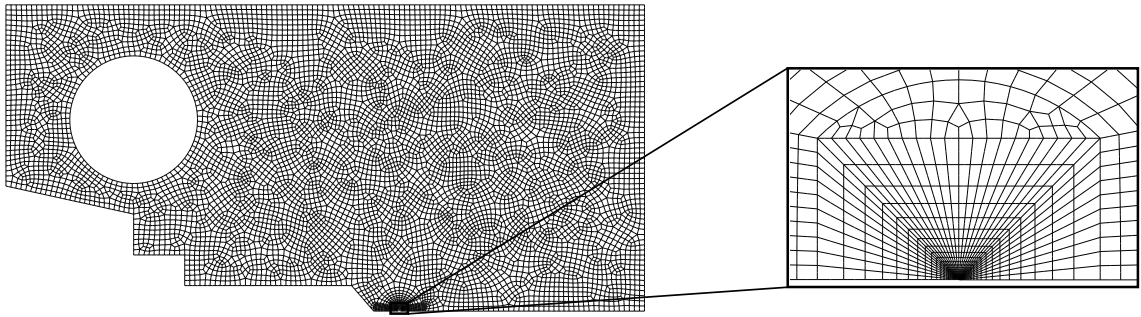}}%
  \caption{General and detailed representation of the finite element mesh employed.}
  \label{fig:Mesh}
\end{figure}

Mimicking the experimental campaign, we model a compact tension specimen of width $W=100$ mm, distance between pins $F=75$ mm and initial notch length $D=51$ mm, referred to as size 2T in \citep{Heerens2002}. The finite element model includes the compact tension specimen and the pins. The load is prescribed by imposing a displacement on the pins, and we model contact between the pins and the specimen by using a surface to surface contact algorithm with finite sliding. The path independent $J$-integral is computed outside of the plastic zone by means of the domain integral method at each load increment. An initial blunting radius of 2 $\mu$m is defined at the crack tip. After a mesh sensitivity analysis, the specimen is discretized with 9800 quadrilateral quadratic plane strain elements. As shown in Fig. \ref{fig:Mesh}, a very refined mesh is used near the crack tip so as to accurately capture the influence of plastic strain gradients.

\section{Results}
\label{Sec:Results}

We begin our analysis by investigating the stress elevation of strain gradient plasticity and its influence on the Weibull stress distribution (Section \ref{Sec:GradientsCrackTipStresses}). Then, we calibrate the Weibull parameters for each temperature and assess the probability of failure due to cleavage with both conventional and MSG plasticity theories (Section \ref{Sec:Cleavage}). Section \ref{Sec:DuctileToBrittle} explores the response across temperatures aiming to gain insight into the role of plastic strain gradients in the ductile-to-brittle transition.\\

\begin{figure}[H]
        \begin{subfigure}[h]{1.1\textwidth}
                \centering
                \includegraphics[scale=1.1]{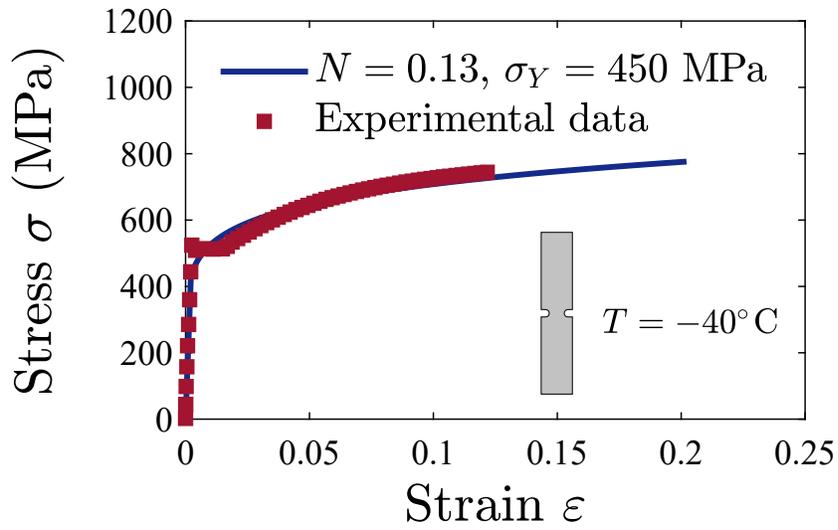}
                \caption{}
                \label{fig:Uniaxial}
        \end{subfigure}\\
		
        \begin{subfigure}[h]{1.1\textwidth}
                \centering
                \includegraphics[scale=1.1]{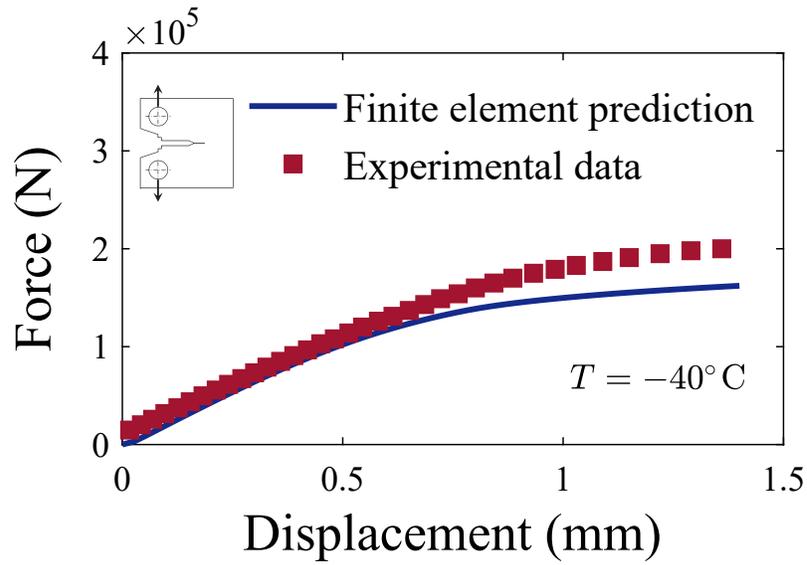}
                \caption{}
                \label{fig:ForceVSdisplacement}
        \end{subfigure}       
        \caption{Calibration of material properties: (a) Uniaxial stress-strain response, and (b) force versus displacement curve in a CT specimen. The case of temperature $T=-40^{\circ}$C is chosen as representative.}\label{fig:Calibration}
\end{figure}

First, we define our uniaxial stress-strain hardening law as
\begin{equation}
\sigma = \sigma_Y \left(1 + \frac{\varepsilon^p}{\sigma_Y}  \right)^N
\end{equation}

\noindent where $N$ is the strain hardening exponent. Thus, in (\ref{EqSflow}), the reference stress equals $\sigma_{ref}=\sigma_Y \left( \frac{E}{\sigma_Y} \right)^N$ and $f \left( \varepsilon^p \right)=\left( \varepsilon^p + \frac{\sigma_Y}{E}\right)^N$. Here, Young's modulus takes the value $E=200$ GPa, and Poisson's ratio equals $\nu=0.3$. We proceed to calibrate $N$ and $\sigma_Y$ with the uniaxial stress-strain data available as part of the Euro toughness data set \citep{Heerens2002}. The values of yield stress $\sigma_Y$ and strain hardening exponent $N$ obtained at each temperature are listed in Table \ref{Tab:MaterialProperties}. A representative fit is shown in Fig. \ref{fig:Uniaxial} for the case of a temperature of $T=-40^\circ$C. As shown in Table \ref{Tab:MaterialProperties}, both $\sigma_Y$ and $N$ decrease with increasing temperature, in agreement with expectations. One should note that the length scale parameter of MSG plasticity has shown a negligible sensitivity to changes in temperature, as measured by \citet{Qian2014} through nanoindentation. Hence, we consider an intermediate value of $\ell=5$ $\mu$m for all temperatures \citep{IJP2016}. \\

\begin{table}[H]
\centering
\caption{Material properties.}
\label{Tab:MaterialProperties}
   {\tabulinesep=1.2mm
   \begin{tabu} {cccccccc}
       \hline
Temperature ($^\circ C$) & -154 &  -91 & -60 & -40 & -20 & 0 & 20 \\ \hline
 Yield stress $\sigma_Y$ (MPa) & 570 & 490 & 470 & 450 & 440 & 430 & 425 \\
 Strain hardening exponent $N$ & 0.14 & 0.14 & 0.13 & 0.13 & 0.13 & 0.12 & 0.12 \\\hline
   \end{tabu}}
\end{table}

The computation of the force versus displacement curves from the calibrated values of $\sigma_Y$ and $N$ shows a good agreement with the experimental data. The results obtained for the representative case of $T=-40^\circ$C are shown in Fig. \ref{fig:ForceVSdisplacement}. The influence of the plastic strain gradients is restricted to a small region next to the crack tip and, consequently, the macroscopic force versus displacement curve is almost insensitive to $\ell$ in the absence of damage. Locally, crack tip stresses are however very sensitive to local strengthening due to gradients of plastic strain.

\subsection{Gradient effects on crack tip stresses}
\label{Sec:GradientsCrackTipStresses}

\begin{figure}[H]
  \makebox[\textwidth][c]{\includegraphics[width=0.9\textwidth]{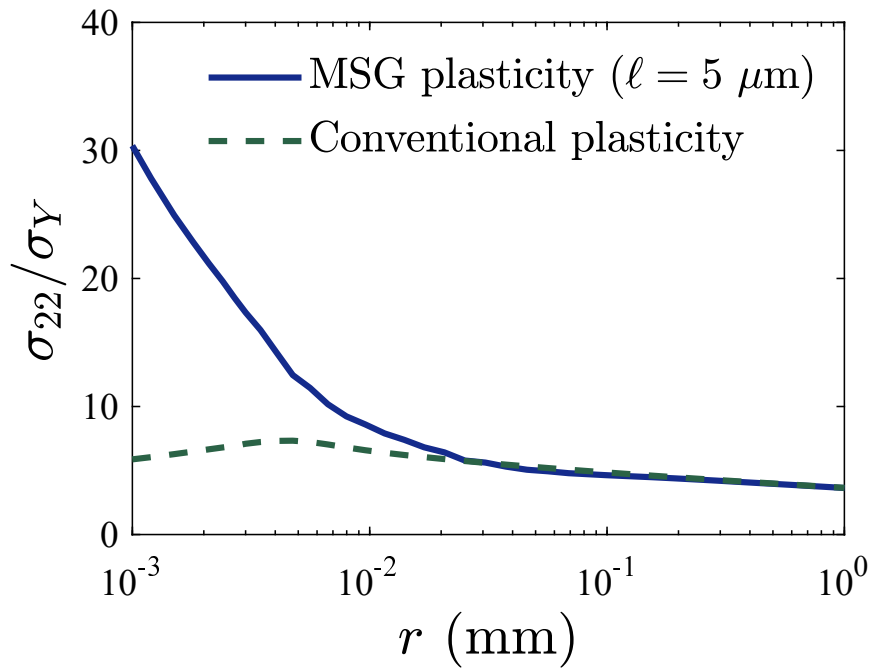}}%
  \caption{Tensile stress along the extended crack plane ($x_2=0$) for both MSG plasticity and conventional plasticity at $J=290$ N/mm. The distance ahead of the crack tip is given in logarithmic scale. The case of temperature $T=-40^{\circ}$C is chosen as representative.}
  \label{fig:CrackTipS22}
\end{figure}

We examine first the tensile stress distribution ahead of the crack for a representative case, $T=-40^{\circ}$C, and a specific load level that falls within the range of critical loads reported in the experiments, $J=290$ N/mm. Results are shown in Fig. \ref{fig:CrackTipS22} for both conventional and MSG plasticity with $\ell=5$ $\mu$m. The tensile stress is normalized by the yield strength of the material at $T=-40^{\circ}$C, and the distance ahead of the crack is shown in logarithmic scale to highlight the different responses given by MSG and conventional plasticity theories. As shown in the figure, far away from the crack tip both MSG plasticity and conventional $J2$ plasticity agree but differences start at about 20-30 $\mu$m ahead of the crack. This distance is sufficiently large to engulf the critical length of various damage mechanisms, including cleavage in ferritic steels. The stress elevation due to plastic strain gradients is associated to large geometrically necessary dislocation (GND) densities that act as obstacles to the motion of statistically stored dislocations and elevate local strength.

\begin{figure}[H]
  \makebox[\textwidth][c]{\includegraphics[width=1.\textwidth]{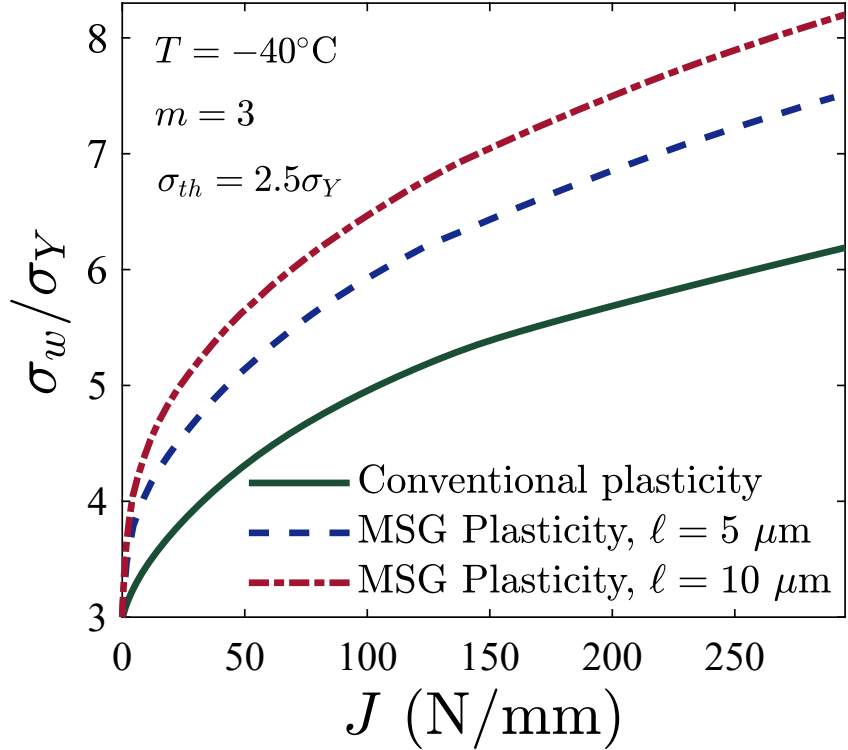}}%
  \caption{Weibull stress dependence on the remote load for both MSG plasticity, with $\ell=5$ $\mu$m and $\ell=10$ $\mu$m, and conventional plasticity. The case of temperature $T=-40^{\circ}$C is chosen as representative.}
  \label{fig:SwvsJ}
\end{figure}

The stress elevation associated with large gradients of plastic strain in the vicinity of a crack influences cleavage models by elevating the Weibull stress $\sigma_w$. We illustrate this by assuming $m=3$ and $\sigma_{th}=2.5\sigma_Y$ and computing the Weibull stress through (\ref{Eq:Sw}) as a function of the remote load. Results are shown in Fig. \ref{fig:SwvsJ} for two representative values of the length scale parameter $\ell=5$ $\mu$m and $\ell=10$ $\mu$m, as well as for conventional plasticity. As shown in the figure, the Weibull stress $\sigma_w$ increases with increasing $\ell$ and differences increase with the remote load. As we shall show below, differences are sensitive to the values of $m$ and $\sigma_{th}$, and the gradient-enhanced $\sigma_w$ elevation can be substantial. Note that, following (\ref{Eq:Pf}), strain gradient plasticity elevates the local probability of failure for a fixed value of $\sigma_u$, $\sigma_{th}$ and $m$.

\subsection{Statistical analysis of cleavage}
\label{Sec:Cleavage}

\begin{figure}[H]
\makebox[\linewidth][c]{%
        \begin{subfigure}[b]{0.55\textwidth}
                \centering
                \includegraphics[scale=0.75]{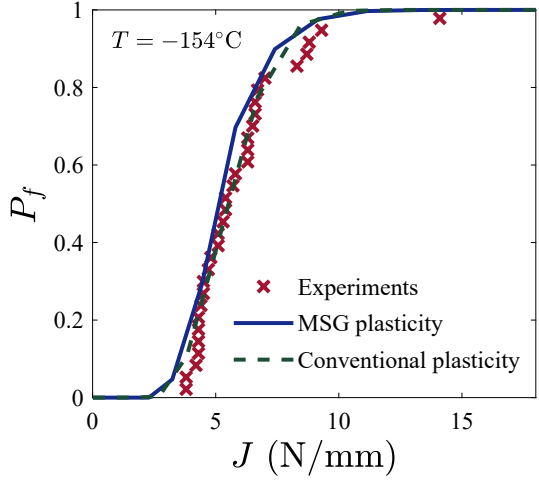}
                \caption{}
                \label{fig:T154}
        \end{subfigure}
        \begin{subfigure}[b]{0.55\textwidth}
                \raggedleft
                \includegraphics[scale=0.75]{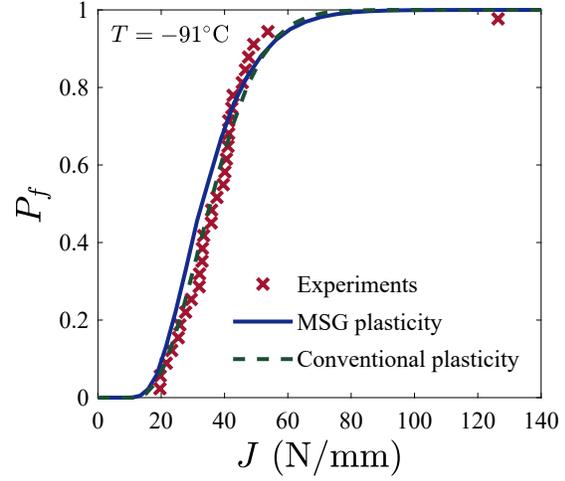}
                \caption{}
                \label{fig:T91}
        \end{subfigure}}

\makebox[\linewidth][c]{%
        \begin{subfigure}[b]{0.55\textwidth}
                \centering
                \includegraphics[scale=0.75]{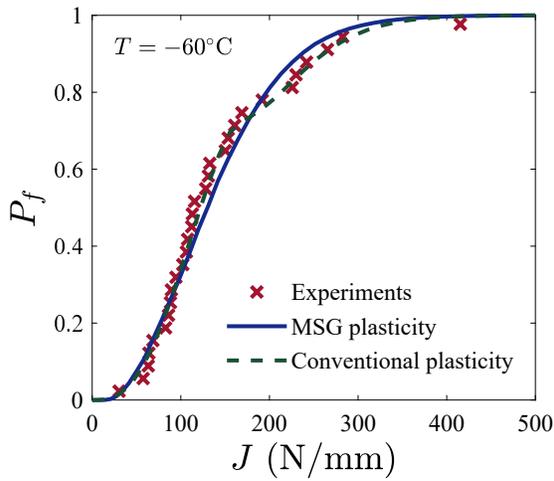}
                \caption{}
                \label{fig:T60}
        \end{subfigure}
        \begin{subfigure}[b]{0.55\textwidth}
                \raggedleft
                \includegraphics[scale=0.75]{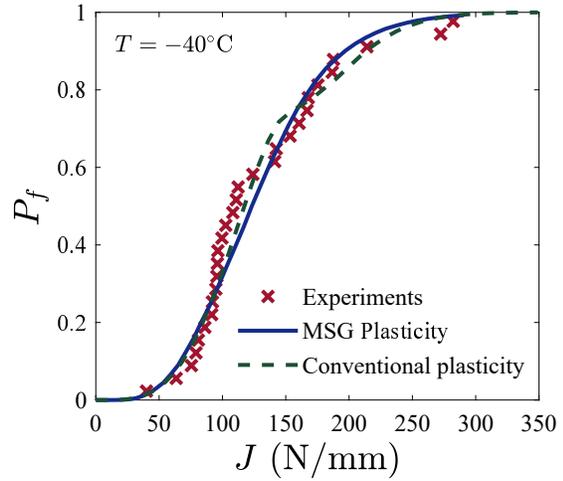}
                \caption{}
                \label{fig:T40}
        \end{subfigure}
        }                
        \caption{Failure probability as a function of the external load. The figure includes the experimental data for 22NiMoCr37 steel \citep{Heerens2002} and the predictons from the present statistical model for the values of $\sigma_{th}$, $\sigma_u$ and $m$ listed in Table \ref{Tab:WeibullParameters}. Temperatures (a) $T=-154^{\circ}$C, (b) $T=-91^{\circ}$C, (c) $T=-60^{\circ}$C and (d) $T=-40^{\circ}$C are chosen as representative.}\label{fig:PfT}
\end{figure}

The statistical framework outlined in Section \ref{Sec:NumModel} is now employed to estimate the probability of failure as a function of the remote load, as quantified by $J$. Fig. \ref{fig:PfT} shows the results obtained for 4 representative temperatures, $T=-154^{\circ}$C, $T=-91^{\circ}$C, $T=-60^{\circ}$C and $T=-40^{\circ}$C. The figure shows the experimental predictions, as given by (\ref{Eq:Pfnj}), along with the results for MSG plasticity with $\ell=5$ $\mu$m and conventional J2 plasticity. Both conventional and MSG plasticity predictions exhibit good agreement with the experiments for the calibrated Weibull parameters.\\

The calibrated values of the modulus $m$, the threshold stress for crack growth $\sigma_{th}$, and the scaling parameter $\sigma_u$ are shown in Table \ref{Tab:WeibullParameters}. Results are shown for 7 temperatures and both strain gradient and conventional plasticity. Considering the effect of plastic strain gradients leads to very significant differences in the values of the calibrated Weibull parameters. Differences between conventional and MSG plasticity are particularly notable in regards to the stress threshold for crack growth $\sigma_{th}$; much larger stresses are needed to propagate micro-cracks if the influence of GNDs is accounted for. Furthermore, qualitative differences are observed in the dependence of the threshold stress with temperature. While the strain gradient plasticity-based prediction exhibits the natural trend of decreasing $\sigma_{th}$ with decreasing $T$ (the material anticipates a reduced barrier to cleavage), this is not the case for conventional plasticity. A plausible explanation behind the scatter observed lies on the fact that the maximum tensile stress is load-independent in  conventional plasticity \citep{McMeeking1977a}; for lower temperatures, a higher stress level is attained for the same $J$ as $\sigma_Y$ is larger. Contrarily, in strain gradient plasticity, crack tip stresses scale with the remote load \citep{EJMAS2019}.\\

In addition, the conventional plasticity results show noticeably high predictions for $m$ at temperatures -60$^\circ$C and -40$^\circ$C. For these two temperatures, there is a clear change in the shape of the $P_f$ versus $J$ curve for values of $P_f$ close to 0.75. Reducing the tolerance of the convergence criterion or changing the initial estimations of $m$ and $\sigma_{th}$ did not have any influence in the outcome of the statistical fitting procedure. Moreover, very similar results were obtained when repeating the procedure with specimens of different geometry; referred to as 0.5T and 1T in \citep{Heerens2002}. The uniqueness of the Weibull parameters (see, e.g., \citealp{Ruggieri2000}) is addressed by repeating the analysis for four different geometries (0.5, 1T, 2T and 4T) and two temperatures (-91$^\circ$C and -20$^\circ$C). Computations reveal very similar values of $m$, $\sigma_{th}$ and $\sigma_u$ to those shown in Table \ref{Tab:WeibullParameters} for both conventional and strain gradient plasticity. Differences are largest with geometry 4T but remain below 10\% in all cases.

\begin{table}[H]
\centering
\caption{Calibration of Weibull parameters for MSG plasticity and conventional plasticity as a function of the temperature.}
\label{Tab:WeibullParameters}
   {\tabulinesep=1.2mm
   \begin{tabu} {cccccccc}
       \hline
MSG plasticity & &  & &  &  & &  \\ \hline
Temperature ($^\circ C$) & -154 &  -91 & -60 & -40 & -20 & 0 & 20 \\ \hline
 $\sigma_u$ (MPa) & 23.6 & 46.9 & 632.3 & 611.7 & 1060.4 & 183.0 & 16948.0\\
 $\sigma_{th}$ (MPa) & 5489.3 & 7295.1 & 7670.6 & 8136.7 & 8295.9 & 19888.0 & 13516.0 \\ 
 $m$ & 2.0 & 1.9 & 2.9 & 3.1 & 3.2 & 1.7 & 12.71 \\\hline
Conventional plasticity & &  & &  &  & &  \\ \hline
Temperature ($^\circ C$) & -154 &  -91 & -60 & -40 & -20 & 0 & 20 \\ \hline
 $\sigma_u$ (MPa) & 9.2 & 14.9 & 1380.4 & 911.3 & 146.0 & 46.11 & 1731.7 \\
 $\sigma_{th}$ (MPa) & 2251.7 & 2459.0 & 1015.7 & 1477.5 & 2289.1 & 2205.0 & 1474.7 \\ 
 $m$ & 1.9 & 1.8 & 13.5 & 12.8 & 3.2 & 0.78 & 19.87 \\\hline 
   \end{tabu}}
\end{table}

\begin{figure}[H]
  \makebox[\textwidth][c]{\includegraphics[width=0.8\textwidth]{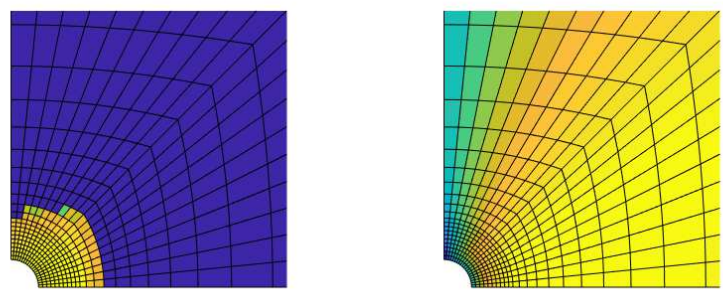}}%
  \caption{Hazard map, showing the local probability of failure in log scale at each material unit for the case of $T=-40^{\circ}$C and $J=282.2$ N/mm.}
  \label{fig:Hazard}
\end{figure}

More insight into the influence of plastic strain gradients on local failure probability can be gained by means of a \emph{hazard map}. In a hazard map, the local probability of failure is shown over the entire engineering component, highlighting the areas that are vulnerable to a specific type of failure \citep{Muniz-Calvente2016}. The local probability of failure is computed as $P_f^i$ in each material unit $i$ from a local $\sigma_w^i$. The results obtained are shown in log scale in Fig. \ref{fig:Hazard} for both conventional and strain gradient plasticity. Important differences can be readily observed. While the local $P_f$ only becomes meaningful close to the crack tip in both cases, the potential damage initiation sites are identified to be much more localized for the case of strain gradient plasticity. In other words, only defects within tens of microns, as opposed to several mm, are identified as fracture-triggering particles when plastic strain gradients are accounted for. The critical distance for cleavage fracture in steels is considered to be significantly smaller than 1 mm \citep{Watanabe1987}.\\
    
\subsection{The ductile to brittle transition}
\label{Sec:DuctileToBrittle}

\begin{figure}[H]
  \makebox[\textwidth][c]{\includegraphics[width=1.2\textwidth]{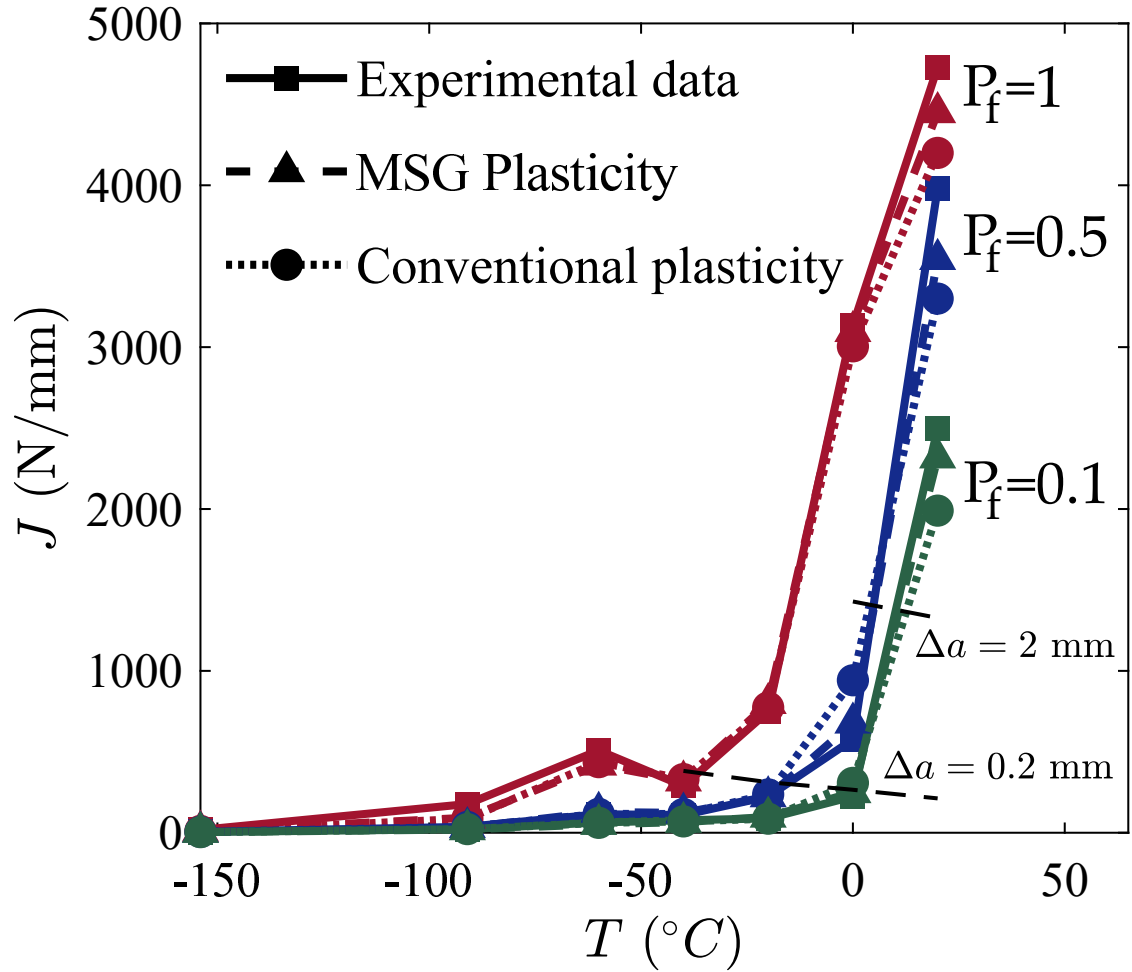}}%
  \caption{Cleavage resistance curves ($P_f=0.5$) and scatter bands ($P_f = 0.1$ and $P_f = 0.9$) for the experimental data, MSG plasticity with $\ell=5$ $\mu$m and conventional plasticity.}
  \label{fig:DuctileToBrittle}
\end{figure}

We then proceed to examine the ductile-to-brittle transition by computing the resistance curves for $P_f=0.1$, 0.5 and 1 across the temperature versus load map. For each value of $P_f$ three curves are shown, the experimental data and the numerically computed results for MSG plasticity and conventional J2 plasticity; see Fig. \ref{fig:DuctileToBrittle}. The results show how the load at which failure is predicted, $P_f=0.5$, increases with the temperature - ductility is enhanced. As the load and the temperature increase, ductile crack growth is observed, with the largest temperatures showing several cases where crack extension equals $\Delta a=2$ mm, the limit value for ductile growth tests \citep{Heerens2002}. In addition, the gradient-enhanced prediction appears to follow more precisely the experimental trend. 

\subsection{Influence of crack tip constraint conditions}
\label{Sec:Tstress}

\begin{figure}[H]
  \makebox[\textwidth][c]{\includegraphics[width=1\textwidth]{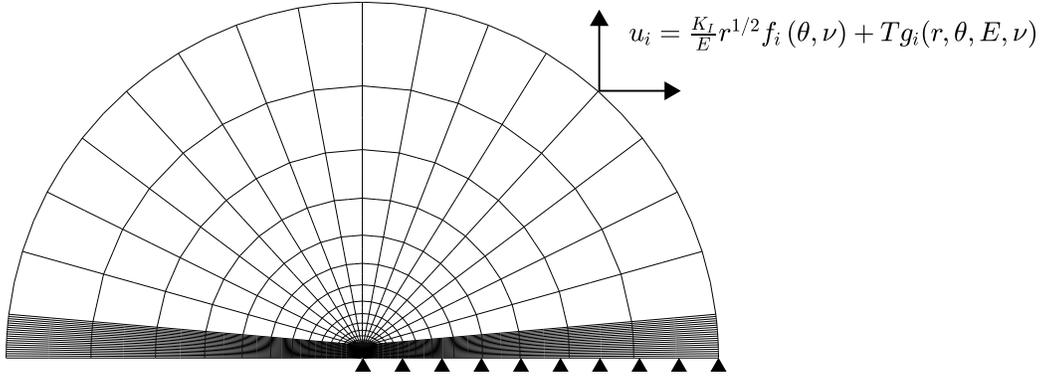}}%
  \caption{Sketch of the modified boundary layer model employed to assess the role of crack tip constraint conditions.}
  \label{fig:ModifiedBoundaryLayer}
\end{figure}

Lastly, we investigate the influence of crack tip constraint conditions by imposing a non-zero elastic T-stress \citep{Betegon1991}. This is achieved by means of the so-called modified boundary layer formulation. Consider a crack plane aligned with the negative $x_1$ axis of the Cartesian reference frame ($x_1$, $x_2$). For a crack tip placed at the origin and a given T value, we choose to prescribe a remote mode I load, $K_I$, by defining the nodal displacements in the outer periphery of the mesh as
\begin{equation}
u_1(r,\theta)=K_I \frac{1+\nu}{E} \sqrt{\frac{r}{2\pi}}cos\left(\frac{\theta}{2}\right)(3-4\nu-cos\theta)+\text{T}\left(\frac{1-\nu^2}{E}\right)r cos\theta
\end{equation}
\begin{equation}
u_2(r,\theta)=K_I \frac{1+\nu}{E} \sqrt{\frac{r}{2\pi}}sin\left(\frac{\theta}{2}\right)(3-4\nu-cos\theta)-\text{T}\left(\frac{\nu(1+\nu)}{E}\right)r sin\theta
\end{equation} 
\noindent where $r$ and $\theta$ are polar coordinates centred at the crack tip. As shown in Fig. \ref{fig:ModifiedBoundaryLayer}, upon exploiting symmetry about the crack plane, only half of the model is analysed. We introduce an initial blunting radius that is 10$^5$ times smaller than the outer radius. The modified boundary layer model is discretized by means of 6422 quadrilateral quadratic plane strain elements.

\begin{figure}[H]
  \makebox[\textwidth][c]{\includegraphics[width=1\textwidth]{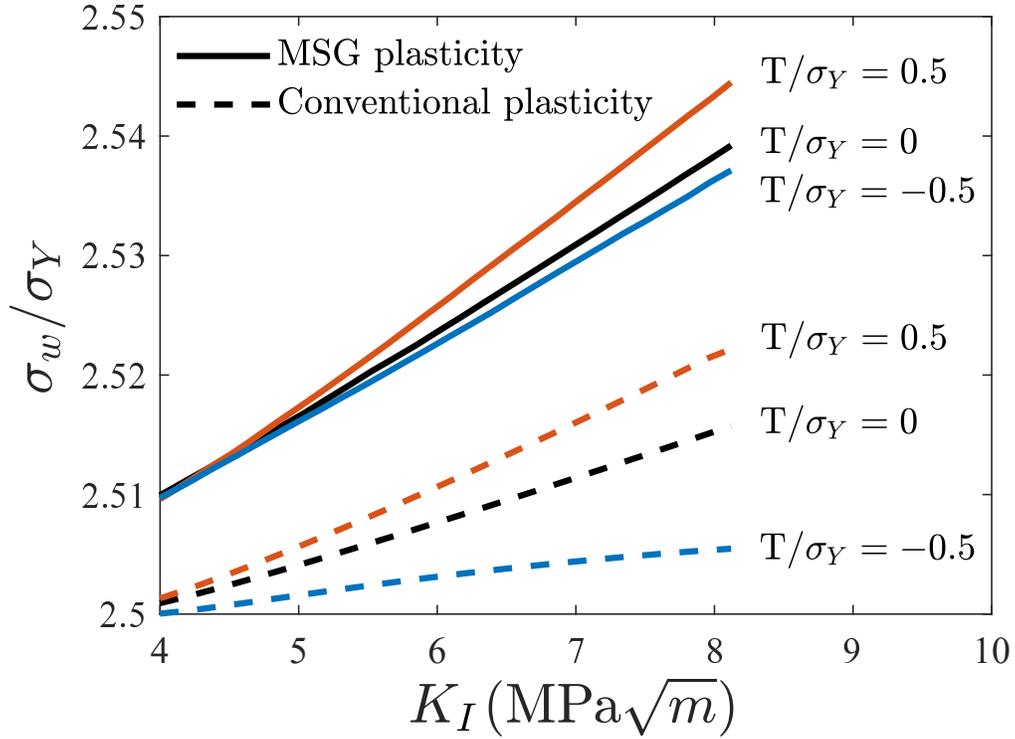}}%
  \caption{Weibull stress dependence on the remote load as a function of the elastic T-stress for conventional plasticity and MSG plasticity, with $\ell=5$ $\mu$m.}
  \label{fig:Tstress}
\end{figure}

The results obtained, in terms of Weibull stress $\sigma_w$ versus remote load $K_I$ are shown in Fig. \ref{fig:Tstress}. As in Fig. \ref{fig:SwvsJ}, we consider a temperature of $T=40^\circ$C and assume $m=3$ and $\sigma_{th}=2.5\sigma_Y$. A range of 3 values of the T-stress is considered: T$/\sigma_Y=-0.5$, T$/\sigma_Y=0$,and  T$/\sigma_Y=0.5$. The same qualitative trends are obtained for both conventional plasticity and strain gradient plasticity; for a given remote load $K_I$, the Weibull stress increases with increasing T. However, conventional plasticity predictions of $\sigma_w$ appear to exhibit a higher sensitivity to crack tip constraint conditions for the values of $m$ and $\sigma_{th}$ assumed.

\section{Conclusions}
\label{Sec:Concluding remarks}

We have presented a 3-parameter statistical framework for cleavage that incorporates the role of large plastic strain gradients in the characterization of crack tip stresses. The model enables to accurately compute Weibull stresses and calibrate - without any prior assumptions - the three statistical parameters: threshold stress $\sigma_{th}$, scaling parameter $\sigma_u$ and modulus $m$. Finite element analysis is used in combination with Weibull statistics to investigate cleavage in ferritic steels with both conventional J2 plasticity and the mechanism-based strain gradient (MSG) plasticity theory. The main findings are:\\

\noindent i) For given values of $\sigma_{th}$, $\sigma_u$ and $m$, strain gradient plasticity effects elevate the Weibull stress and the probability of failure.\\

\noindent ii) The calibrated Weibull parameters for MSG plasticity show significant differences with the values obtained with conventional plasticity. The threshold stress required to trigger cracking in the gradient-enhanced case is 2-8 times larger than its conventional plasticity counterpart.\\

\noindent iii) Hazard maps, where the probability of failure is shown in each material unit, show that defects susceptible of initiating cracking are confined in a much smaller region next to the crack tip in the strain gradient plasticity case.\\

\noindent iv) The probability of failure is computed across the ductile-to-brittle transition, with strain gradient plasticity predictions showing a better agreement with experiments.
 
\section{Acknowledgements}
\label{Acknowledge of funding}

The authors acknowledge valuable discussions with M. Mu\~niz-Calvente (University of Oviedo). The authors would like to acknowledge financial support from the Ministry of Economy and Competitiveness of Spain through grant MAT2014-58738-C3. E. Mart\'{\i}nez-Pa\~neda additionally acknowledges financial support from Wolfson College Cambridge (Junior Research Fellowship) and from the Royal Commission for the 1851 Exhibition through their Research Fellowship programme (RF496/2018).




\bibliographystyle{elsarticle-harv}
\bibliography{library}

\begin{thebibliography}{48}
\expandafter\ifx\csname natexlab\endcsname\relax\def\natexlab#1{#1}\fi
\expandafter\ifx\csname url\endcsname\relax
  \def\url#1{\texttt{#1}}\fi
\expandafter\ifx\csname urlprefix\endcsname\relax\def\urlprefix{URL }\fi

\bibitem[{Aifantis(1984)}]{Aifantis1984}
Aifantis, E.~C., 1984. {On the Microstructural Origin of Certain Inelastic
  Models}. Journal of Engineering Materials and Technology 106~(4), 326.

\bibitem[{Anand et~al.(2005)Anand, Gurtin, Lele, and Gething}]{Anand2005}
Anand, L., Gurtin, M.~E., Lele, S.~P., Gething, C., 2005. {A one-dimensional
  theory of strain-gradient plasticity: Formulation, analysis, numerical
  results}. Journal of the Mechanics and Physics of Solids 53~(8), 1789--1826.

\bibitem[{Ashby(1970)}]{Ashby1970}
Ashby, M.~F., 1970. {The deformation of plastically non-homogeneous materials}.
  Philosophical Magazine 21~(170), 399--424.

\bibitem[{Bagchi and Evans(1996)}]{Bagchi1996}
Bagchi, A., Evans, A.~G., 1996. {The Mechanics and Physics of Thin-Film
  Decohesion and Its Measurement}. Interface Science 3, 169--193.

\bibitem[{Beremin(1983)}]{Beremin1983}
Beremin, F.~M., 1983. {A local criterion for cleavage fracture of a nuclear
  pressure vessel steel}. Metallurgical Transactions A 14~(11), 2277--2287.

\bibitem[{Betegón and Hancock(1991)}]{Betegon1991}
Betegón, C., Hancock, J.~W., 1991. {Two-Parameter Characterization of
  Elastic-Plastic Crack-Tip Fields}. Journal of Applied Mechanics 58~(1),
  104--110.

\bibitem[{Brinckmann and Siegmund(2008)}]{Brinckmann2008}
Brinckmann, S., Siegmund, T., 2008. {Computations of fatigue crack growth with
  strain gradient plasticity and an irreversible cohesive zone model}.
  Engineering Fracture Mechanics 75~(8), 2276--2294.

\bibitem[{Cleveringa et~al.(2000)Cleveringa, {Van der Giessen}, and
  Needleman}]{Cleveringa2000}
Cleveringa, H., {Van der Giessen}, E., Needleman, A., 2000. {A discrete
  dislocation analysis of mode I crack growth}. Journal of the Mechanics and
  Physics of Solids 48~(6–7), 1133--1157.

\bibitem[{Elssner et~al.(1994)Elssner, Korn, and R{\"{u}}hle}]{Elssner1994}
Elssner, G., Korn, D., R{\"{u}}hle, M., 1994. {The influence of interface
  impurities on fracture energy of UHV diffusion bonded metal-ceramic
  bicrystals}. Scripta Metallurgica et Materiala 31~(8), 1037--1042.

\bibitem[{Fleck and Hutchinson(2001)}]{Fleck2001}
Fleck, N.~A., Hutchinson, J.~W., 2001. {A reformulation of strain gradient
  plasticity}. Journal of the Mechanics and Physics of Solids 49~(10),
  2245--2271.

\bibitem[{Fleck et~al.(1994)Fleck, Muller, Ashby, and Hutchinson}]{Fleck1994}
Fleck, N.~A., Muller, G.~M., Ashby, M.~F., Hutchinson, J.~W., 1994. {Strain
  gradient plasticity: Theory and Experiment}. Acta Metallurgica et Materialia
  42~(2), 475--487.

\bibitem[{Gao et~al.(1999)Gao, Hang, Nix, and Hutchinson}]{Gao1999}
Gao, H., Hang, Y., Nix, W.~D., Hutchinson, J.~W., 1999. {Mechanism-based strain
  gradient plasticity - I. Theory}. Journal of the Mechanics and Physics of
  Solids 47~(6), 1239--1263.

\bibitem[{Gil-Sevillano(2001)}]{Sevillano2001}
Gil-Sevillano, J., 2001. {The effective threshold for fatigue crack
  propagation: A plastic size effect?} Scripta Materialia 44~(11), 2661--2665.

\bibitem[{Heerens and Hellmann(2002)}]{Heerens2002}
Heerens, J., Hellmann, D., 2002. {Development of the Euro fracture toughness
  dataset}. Engineering Fracture Mechanics 69~(4), 421--449.

\bibitem[{Huang et~al.(2004)Huang, Qu, Hwang, Li, Gao, Huang, Qu, Hwang, Li,
  and Gao}]{Huang2004a}
Huang, Y., Qu, S., Hwang, K.~C., Li, M., Gao, H., Huang, Y., Qu, S., Hwang,
  K.~C., Li, M., Gao, H., 2004. {A conventional theory of mechanism-based
  strain gradient plasticity}. International Journal of Plasticity 20~(4-5),
  753--782.

\bibitem[{Hutchinson(1997)}]{Hutchinson1997}
Hutchinson, J.~W., 1997. {Linking scales in fracture mechanics}. Advances in
  Fracture Research, Proceedings of ICF10, 1 -- 14.

\bibitem[{Irani et~al.(2017)Irani, Remmers, and Deshpande}]{Irani2017}
Irani, N., Remmers, J. J.~C., Deshpande, V.~S., 2017. {A discrete dislocation
  analysis of hydrogen-assisted mode-I fracture}. Mechanics of Materials 105,
  67--79.

\bibitem[{Jiang et~al.(2001)Jiang, Huang, Zhuang, and Hwang}]{Jiang2001}
Jiang, H., Huang, Y., Zhuang, Z., Hwang, K.~C., 2001. {Fracture in
  mechanism-based strain gradient plasticity}. Journal of the Mechanics and
  Physics of Solids 49~(5), 979--993.

\bibitem[{Komaragiri et~al.(2008)Komaragiri, Agnew, Gangloff, and
  Begley}]{Komaragiri2008}
Komaragiri, U., Agnew, S.~R., Gangloff, R.~P., Begley, M.~R., 2008. {The role
  of macroscopic hardening and individual length-scales on crack tip stress
  elevation from phenomenological strain gradient plasticity}. Journal of the
  Mechanics and Physics of Solids 56~(12), 3527--3540.

\bibitem[{Korn et~al.(2002)Korn, Elssner, Cannon, and Ruhle}]{Korn2002}
Korn, D., Elssner, G., Cannon, R.~M., Ruhle, M., 2002. {Fracture properties of
  interfacially doped Nb-A12O3 bicrystals: I, fracture characteristics}. Acta
  Materialia 50~(15), 3881--3901.

\bibitem[{Mart{\'{i}}nez-Pa{\~{n}}eda and Beteg{\'{o}}n(2015)}]{IJSS2015}
Mart{\'{i}}nez-Pa{\~{n}}eda, E., Beteg{\'{o}}n, C., 2015. {Modeling damage and
  fracture within strain-gradient plasticity}. International Journal of Solids
  and Structures 59, 208--215.

\bibitem[{Mart{\'{i}}nez-Pa{\~{n}}eda
  et~al.(2017{\natexlab{a}})Mart{\'{i}}nez-Pa{\~{n}}eda, del Busto, and
  Beteg{\'{o}}n}]{TAFM2017}
Mart{\'{i}}nez-Pa{\~{n}}eda, E., del Busto, S., Beteg{\'{o}}n, C.,
  2017{\natexlab{a}}. {Non-local plasticity effects on notch fracture
  mechanics}. Theoretical and Applied Fracture Mechanics 92, 276--287.

\bibitem[{Mart{\'{i}}nez-Pa{\~{n}}eda
  et~al.(2016{\natexlab{a}})Mart{\'{i}}nez-Pa{\~{n}}eda, del Busto, Niordson,
  and Beteg{\'{o}}n}]{IJHE2016}
Mart{\'{i}}nez-Pa{\~{n}}eda, E., del Busto, S., Niordson, C.~F., Beteg{\'{o}}n,
  C., 2016{\natexlab{a}}. {Strain gradient plasticity modeling of hydrogen
  diffusion to the crack tip}. International Journal of Hydrogen Energy
  41~(24), 10265--10274.

\bibitem[{Mart{\'{i}}nez-Pa{\~{n}}eda et~al.(2019)Mart{\'{i}}nez-Pa{\~{n}}eda,
  Deshpande, Niordson, and Fleck}]{JMPS2019}
Mart{\'{i}}nez-Pa{\~{n}}eda, E., Deshpande, V.~S., Niordson, C.~F., Fleck,
  N.~A., 2019. {The role of plastic strain gradients in the crack growth
  resistance of metals}. Journal of the Mechanics and Physics of Solids 126,
  136--150.

\bibitem[{Mart{\'{i}}nez-Pa{\~{n}}eda and Fleck(2019)}]{EJMAS2019}
Mart{\'{i}}nez-Pa{\~{n}}eda, E., Fleck, N.~A., 2019. {Mode I crack tip fields:
  Strain gradient plasticity theory versus J2 flow theory}. European Journal of
  Mechanics - A/Solids 75, 381--388.

\bibitem[{Mart{\'{i}}nez-Pa{\~{n}}eda
  et~al.(2017{\natexlab{b}})Mart{\'{i}}nez-Pa{\~{n}}eda, Natarajan, and
  Bordas}]{CM2017}
Mart{\'{i}}nez-Pa{\~{n}}eda, E., Natarajan, S., Bordas, S., 2017{\natexlab{b}}.
  {Gradient plasticity crack tip characterization by means of the extended
  finite element method}. Computational Mechanics 59, 831--842.

\bibitem[{Mart{\'{i}}nez-Pa{\~{n}}eda and Niordson(2016)}]{IJP2016}
Mart{\'{i}}nez-Pa{\~{n}}eda, E., Niordson, C.~F., 2016. {On fracture in finite
  strain gradient plasticity}. International Journal of Plasticity 80,
  154--167.

\bibitem[{Mart{\'{i}}nez-Pa{\~{n}}eda
  et~al.(2016{\natexlab{b}})Mart{\'{i}}nez-Pa{\~{n}}eda, Niordson, and
  Gangloff}]{AM2016}
Mart{\'{i}}nez-Pa{\~{n}}eda, E., Niordson, C.~F., Gangloff, R.~P.,
  2016{\natexlab{b}}. {Strain gradient plasticity-based modeling of hydrogen
  environment assisted cracking}. Acta Materialia 117, 321--332.

\bibitem[{McMeeking(1977)}]{McMeeking1977a}
McMeeking, R.~M., 1977. {Finite deformation analysis of crack-tip opening in
  elastic-plastic materials and implications for fracture}. Journal of the
  Mechanics and Physics of Solids 25~(5), 357--381.

\bibitem[{Mu{\~{n}}iz-Calvente et~al.(2015)Mu{\~{n}}iz-Calvente,
  {Fern{\'{a}}ndez Canteli}, Shlyannikov, and Castillo}]{Muniz-Calvente2015}
Mu{\~{n}}iz-Calvente, M., {Fern{\'{a}}ndez Canteli}, A., Shlyannikov, V.,
  Castillo, E., 2015. {Probabilistic Weibull Methodology for Fracture
  Prediction of Brittle and Ductile Materials}. Applied Mechanics and Materials
  784~(2), 443--451.

\bibitem[{Muniz-Calvente et~al.(2016)Muniz-Calvente, Ramos, Pelayo, Lamela, and
  Fern{\'{a}}ndez-Canteli}]{Muniz-Calvente2016a}
Muniz-Calvente, M., Ramos, A., Pelayo, F., Lamela, M.~J.,
  Fern{\'{a}}ndez-Canteli, A., 2016. {Statistical joint evaluation of fracture
  results from distinct experimental programs: An application to annealed
  glass}. Theoretical and Applied Fracture Mechanics 85, 149--157.

\bibitem[{Mu{\~{n}}iz-Calvente et~al.(2016)Mu{\~{n}}iz-Calvente, Ramos,
  Shlyannikov, Lamela, and Fern{\'{a}}ndez-Canteli}]{Muniz-Calvente2016}
Mu{\~{n}}iz-Calvente, M., Ramos, A., Shlyannikov, V., Lamela, M.~J.,
  Fern{\'{a}}ndez-Canteli, A., 2016. {Hazard maps and global probability as a
  way to transfer standard fracture results to reliable design of real
  components}. Engineering Failure Analysis 69, 135--146.

\bibitem[{Nix and Gao(1998)}]{Nix1998}
Nix, W.~D., Gao, H.~J., 1998. {Indentation size effects in crystalline
  materials: A law for strain gradient plasticity}. Journal of the Mechanics
  and Physics of Solids 46~(3), 411--425.

\bibitem[{Papazafeiropoulos et~al.(2017)Papazafeiropoulos,
  Mu{\~{n}}iz-Calvente, and Mart{\'{i}}nez-Pa{\~{n}}eda}]{AES2017}
Papazafeiropoulos, G., Mu{\~{n}}iz-Calvente, M., Mart{\'{i}}nez-Pa{\~{n}}eda,
  E., 2017. {Abaqus2Matlab: A suitable tool for finite element
  post-processing}. Advances in Engineering Software 105, 9--16.

\bibitem[{Pineau et~al.(2016)Pineau, Benzerga, and Pardoen}]{Pineau2016}
Pineau, A., Benzerga, A.~A., Pardoen, T., 2016. {Failure of metals I: Brittle
  and ductile fracture}. Acta Materialia 107, 424--483.

\bibitem[{Pribe et~al.(2019)Pribe, Siegmund, Tomar, and Kruzic}]{Pribe2019}
Pribe, J.~D., Siegmund, T., Tomar, V., Kruzic, J.~J., 2019. {Plastic strain
  gradients and transient fatigue crack growth: a computational study}.
  International Journal of Fatigue 120, 283--293.

\bibitem[{Qian et~al.(2011)Qian, Zhang, and Swaddiwudhipong}]{Qian2011}
Qian, X., Zhang, S., Swaddiwudhipong, S., 2011. {Calibration of Weibull
  parameters using the conventional mechanism-based strain gradient
  plasticity}. Engineering Fracture Mechanics 78~(9), 1928--1944.

\bibitem[{Qian et~al.(2014)Qian, Zhang, Swaddiwudhipong, and Shen}]{Qian2014}
Qian, X., Zhang, S., Swaddiwudhipong, S., Shen, L., 2014. {Temperature
  dependence of material length scale for strain gradient plasticity and its
  effect on near-tip opening displacement}. Fatigue and Fracture of Engineering
  Materials and Structures 37~(2), 157--170.

\bibitem[{Qiu et~al.(2003)Qiu, Huang, Wei, Gao, and Hwang}]{Qiu2003}
Qiu, X., Huang, Y., Wei, Y., Gao, H., Hwang, K., 2003. {The flow theory of
  mechanism-based strain gradient plasticity}. Mechanics of Materials 35~(3-6),
  245--258.

\bibitem[{Raynolds et~al.(1996)Raynolds, Smith, Zhao, and
  Srolovitz}]{Raynolds1996}
Raynolds, J.~E., Smith, J.~R., Zhao, G.~L., Srolovitz, D.~J., 1996. {Adhesion
  in NiAl-Cr from first principles}. Physical Review B 53~(20), 13883--13890.

\bibitem[{Ruggieri et~al.(2000)Ruggieri, Gao, and Dodds}]{Ruggieri2000}
Ruggieri, C., Gao, X., Dodds, R.~H., 2000. {Transferability of elastic-plastic
  fracture toughness using the Weibull stress approach: Significance of
  parameter calibration}. Engineering Fracture Mechanics 67~(2), 101--117.

\bibitem[{Shi et~al.(2001)Shi, Huang, Jiang, Hwang, and Li}]{Shi2001}
Shi, M., Huang, Y., Jiang, H., Hwang, K.~C., Li, M., 2001. {The boundary-layer
  effect on the crack tip field in mechanism-based strain gradient plasticity}.
  International Journal of Fracture 112~(1), 23--41.

\bibitem[{St{\"{o}}lken and Evans(1998)}]{Stolken1998}
St{\"{o}}lken, J.~S., Evans, A.~G., 1998. {A microbend test method for
  measuring the plasticity length scale}. Acta Materialia 46~(14), 5109--5115.

\bibitem[{Suo et~al.(1993)Suo, Shih, and Varias}]{Suo1993}
Suo, Z., Shih, C.~F., Varias, A.~G., 1993. {A theory for cleavage cracking in
  the presence of plastic flow}. Acta Metallurgica Et Materialia 41~(5),
  1551--1557.

\bibitem[{Tvergaard and Hutchinson(1992)}]{Tvergaard1992}
Tvergaard, V., Hutchinson, J.~W., 1992. {The relation between crack growth
  resistance and fracture process parameters in elastic-plastic solids}.
  Journal of the Mechanics and Physics of Solids 40~(6), 1377--1397.

\bibitem[{Tvergaard and Niordson(2008)}]{Tvergaard2008}
Tvergaard, V., Niordson, C.~F., 2008. {Size effects at a crack-tip interacting
  with a number of voids}. Philosophical Magazine 88~(30-32), 3827--3840.

\bibitem[{Watanabe et~al.(1987)Watanabe, Iwadate, Tanaka, Yokobori, and
  Ando}]{Watanabe1987}
Watanabe, J., Iwadate, T., Tanaka, Y., Yokobori, T., Ando, K., 1987. {Fracture
  toughness in the transition region}. Engineering Fracture Mechanics 28~(5-6),
  589--600.

\bibitem[{Wei(2006)}]{Wei2006}
Wei, Y., 2006. {A new finite element method for strain gradient theories and
  applications to fracture analyses}. European Journal of Mechanics, A/Solids
  25~(6), 897--913.

\end{thebibliography}


\end{document}